\documentstyle[preprint,aps]{revtex}

\begin{document}
\title{Inelastic Coulomb scattering rates due to acoustic and optical plasmon modes
in coupled quantum wires.}
\author{Marcos R. S. Tavares and Guo-Qiang Hai}
\address{Instituto de F\'{\i}sica de S\~{a}o Carlos, Universidade de S\~{a}o Paulo,\\
13560-970 S\~{a}o Carlos, SP, Brazil.}
\maketitle

\begin{abstract}
We report a theoretical study on the inelastic Coulomb scattering rate of an
injected electron in two coupled quantum wires\ in quasi-one-dimensional
doped semiconductors. Two peaks appear in the scattering spectrum due to the
optical and the acoustic plasmon scattering in the system. We find that the
scattering rate due to the optical plasmon mode is similar to that in a
single wire but the acoustic plasmon scattering depends crucially on its
dispersion relation at small $q$. Furthermore, the effects of tunneling
between the two wires are studied on the inelastic Coulomb scattering rate.
We show that a weak tunneling can strongly affect the acoustic plasmon
scattering.
\end{abstract}


\section{Introduction}

Recently, single-particle properties of electrons in quasi-one-dimensional
(Q1D) electron systems have attracted considerable interest. With the
theoretical calculations of quasiparticle renormalization factor\cite{mhbook}
and the momentum distribution function around the Fermi surface, Hu and Das
Sarma\cite{sarmagrande} have clarified that a clean 1D electron system shows
the Luttinger liquid behavior, but even a slightest amount of impurities
restores the Fermi surface and the Fermi-liquid behavior remains. Within a
one-subband model, they evaluated the self-energy due to electron-electron
Coulomb interaction in unclean Q1D systems by using the leading-order GW
dynamical screening approximation.\cite{mhbook,quinn} Within such an
approximation, Hwang and Das Sarma\cite{hwang} obtained the band-gap
renormalization in photoexcited doped-semiconductor quantum wire in the
presence of plasmon-phonon coupling. In particular, the inelastic Coulomb
scattering rate plays an important role in relaxation processes of an
injected \ electron in the conduction band. The lifetime of the injected
electron, determined by this scattering rate, can be measured by femtosecond
time-resolved photoemission spectroscopy.\cite{Xuetal} The relaxation
processes of an injected electron occur through the scattering channels due
to different excitations in the system, such as quasiparticle excitations,
plasmons, and phonons.\cite{zheng,muitos} Its lifetime provides information
on the interactions between the electron and the different excitations. The
relaxation mechanism is important because of its technological relevance, as
most semiconductor-based devices operate under high-field and hot-electron
conditions.\cite{doisdozheng}

On the other hand, tunneling effects have provide new devices formed by
coupled Q1D doped semiconductors\cite{exp1d} and attracted considerable
theoretical interest because of their fundamental applicability. In this
work we present a theoretical study on the inelastic Coulomb scattering
rates in coupled bi-wire electron gas systems. A particular attention will
be devoted to the effects of weak resonant tunneling. We find that a weak
resonance tunneling can introduce a strong intersubband inelastic Coulomb
scattering by emitting an acoustic plasmon. The emission of optical plasmon,
on the other hand, is provided by intrasubband scattering of injected
electrons.

The rest of the paper is organized as follows. In Sec. II, we present the
theoretical formalism of the inelastic Coulomb scattering rates in a
multisubband Q1D system of coupled quantum wires. Sec. III is devoted to
analyze the inelastic-scattering rates for a bi-wire system in the absence
of tunneling between the wires. As an extension of such calculations we show
in Sec. IV the numerical results in the presence of weak resonant tunneling.
Finally, we summarize our results in Sec. V.

\section{Theoretical formulation}

We consider a two-dimensional system in the $xy$ plane subjected to an
additional confinement in the $y$-direction which forms two quantum wires
parallel to each other in the $x$-direction. The confinement potential in
the $y$-direction is taken to be of square well type of barrier height $%
V_{0} $ and well widths $W_{1}$ and $W_{2}$ representing the first and the
second wire, respectively. The potential barrier between the two wires is of
width$\ W_{b}$. The subband energies $E_{n}$ and the wave functions $\phi
_{n}(y)$ are obtained from the numerical solution of the one-dimensional
Schr\"{o}dinger equation in the $y$-direction. We restrict ourselves to the
case where $n=1,2$ and define $\omega _{0}=E_{2}-E_{1}$ as being the gap
between the two subbands. The interpretation of the index $n$\thinspace
depends on tunneling between the two wires. \ When there is no tunneling,
the wavefunction $\phi _{n}(y)$ of the subband $E_{n}$ is localized in
quantum wire $n$. Clearly, it is wire index. For two symmetric quantum
wires, i.e., $W_{1}=W_{2},$ one has $E_{2}=E_{1}$ or $\omega _{0}=0.$ When
tunneling occurs, the wavefunction of each subband spreads in two quantum
wires. In this case, $n$ is interpreted as subband index. For two symmetric
quantum wires with tunneling, the wavefunctions of the two lowest
eigenstates are symmetric and antisymmetric. In this case, the two wires are
in the resonant tunneling condition and the gap between the two subbands is
denoted by $\Delta _{SAS}=\omega _{0}$.

In a multisubband Q1D system, the inelastic Coulomb scattering rate for an
injected electron in subband $n$ with momentum $k$ can be obtained by the
imaginary part of the screened exchange self-energy $\Sigma {}_{n}\left[
k,\xi _{n}(k)\right] $,\cite{mhbook} where, $\xi _{n}\left( k\right) =\hbar
^{2}k^{2}/2m^{\ast }+E_{n}-E_{F}$ is the electron energy with respect to the
Fermi energy $E_{F}$ and $m^{\ast }$ the electron effective mass. At zero
temperature, this self-energy can be obtained from the leading terms of the
Dyson's equation for the dressed electron Green's function,\cite{vinter} and
given by 
\begin{equation}
\Sigma {}_{n}\left[ k,\xi _{n}(k)\right] =\frac{i}{(2\pi )^{2}}\int dq\int
d\omega ^{\prime }\sum_{n_{1}}V_{nn_{1}n_{1}n}^{s}(q,\omega ^{\prime
})G_{n_{1}}^{(0)}\left( k+q,\xi _{n}(k)-\omega ^{\prime }\right) ,
\label{self1}
\end{equation}
where $G_{n_{1}}^{(0)}\left( k,\omega \right) $ is the Greens function of
noninteracting electrons and $V_{nn_{1}n_{1}n}^{s}(q,\omega )$ is the
dynamically screened electron-electron Coulomb potential. The screened
Coulomb potential is related to the dielectric function $\varepsilon
_{nn^{\prime }mm^{\prime }}(q,\omega )$ and the bare electron-electron
interaction potential $V_{nn^{\prime }mm^{\prime }}(q)$ through the equation 
\begin{equation}
\sum_{ll^{\prime }}\epsilon _{ll^{\prime }nn^{\prime }}(q,\omega
)V_{ll^{\prime }mm^{\prime }}^{s}(q,\omega )=V_{nn^{\prime }mm^{\prime }}(q).
\label{diel}
\end{equation}
Similarly to the one-band model\cite{sarmagrande}, the self-energy in Eq.\ (%
\ref{self1}) can be separated into the frequency-independent exchange and
the correlation part, $\Sigma _{n}\left[ k,\xi _{n}(k)\right] =\Sigma
_{n}^{ex}(k)+\Sigma _{n}^{cor}\left[ k,\xi _{n}(k)\right] .$ The exchange
part is given by 
\begin{equation}
\Sigma _{n}^{ex}(k)=-\frac{1}{2\pi }\int
dq\sum_{n_{1}}V_{nn_{1}n_{1}n}(q)\;f_{n_{1}}\left( \xi _{n_{1}}(k+q)\right) ,
\end{equation}
where $f_{n}\left( \xi \right) $ is the Fermi-Dirac distribution function.
Notice that $\Sigma {}_{n}^{ex}(k)$ is real because the bare
electron-electron Coulomb potential $V_{nn_{1}n_{1}n}(q)$ is totally real.
Therefore, one only needs to analyze the imaginary part of $\Sigma _{n}^{cor}%
\left[ k,\xi _{n}(k)\right] $, since it gives rise to the imaginary part of
the self-energy which we are interested in. After some algebra, we find that
the Coulomb inelastic-scattering rate for an electron in a subband $n$ with
momentum $k$ is given by 
\begin{equation}
\sigma _{n}(k)=-%
\mathop{\rm Im}%
\Sigma {}_{n}^{cor}\left[ k,\xi _{n}(k)\right] =\sum_{n^{\prime }}\sigma
_{n,n^{\prime }}(k),  \label{praver}
\end{equation}
with 
\[
\sigma _{n,n^{\prime }}(k)=\frac{1}{2\pi }\int dq%
\mathop{\rm Im}%
\left\{ V_{nn^{\prime }n^{\prime }n}^{s}[q,\xi _{n^{\prime }}(k+q)-\xi
_{n}(k)]\right\} 
\]
\begin{equation}
\times \left\{ \theta \left( \xi _{n}(k)-\xi _{n^{\prime }}(k+q)\right)
-\theta \left( -\xi _{n^{\prime }}(k+q)\right) \right\} ,  \label{sigma}
\end{equation}
where $\theta \left( x\right) $ is the standard step function. In the above
equation, the frequency integration has already been carried out, since the
bare Green's function $G_{n_{1}}^{(0)}$ can be written as a Dirac delta
function of $\omega $.

For the present coupled quantum wire systems with two occupied subbands, \
the multisubband dielectric function within the random-phase approximation
(RPA) is given by 
\begin{equation}
\varepsilon _{nn^{\prime }mm^{\prime }}(q,\omega )=\delta _{nm}\delta
_{n^{\prime }m^{\prime }}-\Pi _{nn^{\prime }}(q,\omega )V_{nn^{\prime
}mm^{\prime }}(q).  \label{diel1}
\end{equation}
The function $\Pi _{nn^{\prime }}(q,\omega )$ is the 1D non-interacting
irreducible polarizability at zero temperature for a system free from any
impurity scattering. In the presence of impurity scattering, we use Mermin's
formula \cite{Mermin} 
\begin{equation}
\Pi _{nn^{\prime }}^{\gamma }(q,\omega )=\frac{(\omega +i\gamma )\Pi
_{nn^{\prime }}(q,\omega +i\gamma )}{\omega +i\gamma \left[ \Pi _{nn^{\prime
}}(q,\omega +i\gamma )/\Pi _{nn^{\prime }}(q,0)\right] }  \label{pol2}
\end{equation}
to obtain the polarizability including the effect of level broadening
through a phenomenological damping constant $\gamma $. The Coulomb potential 
\[
V_{nn^{\prime }mm^{\prime }}(q)=\frac{2e^{2}}{\epsilon _{0}}\int dy\int
dy^{\prime }\phi _{n}(y)\phi _{n^{\prime }}(y)K_{0}\left( q\left|
y-y^{\prime }\right| \right) \phi _{m}(y^{\prime })\phi _{m^{\prime
}}(y^{\prime }) 
\]
is calculated by using the numerical solution of the electron wavefunction $%
\phi _{n}(y).$ Here, $\epsilon _{0}$ is the static lattice dielectric
constant, $e$ is the electron charge, and $K_{0}\left( q\left| y-y^{\prime
}\right| \right) $ is the zeroth-order modified Bessel function of the
second order. The electron-electron Coulomb interaction describes
two-particle scattering events. We observe the following characteristics of
the electron-electron Coulomb interaction in the coupled quantum wires
representing different physical scattering processes: $V_{1111}(q)=V_{A},$ $%
V_{2222}(q)=V_{B},$ and $V_{1122}(q)=V_{2211}(q)=V_{C}$ represent\ the
scattering in which the electrons keep in their original wires or subbands; $%
V_{1212}(q)=V_{2121}(q)=V_{1221}(q)=V_{2112}(q)=V_{D}$ represent the
scattering in which both electrons change their wire or subband indices; $%
V_{1112}(q)=V_{1121}(q)=V_{1211}(q)=V_{2111}(q)=V_{J}$ and $%
V_{2212}(q)=V_{2221}(q)=V_{1222}(q)=V_{2122}(q)=$ $V_{H}$ indicating the
scattering in which only one of the electrons suffers the interwire or
intersubband transition.\ When there is no tunneling, $V_{D}=V_{H}=V_{J}=0$.
Clearly, they are responsible for tunneling effects. We also notice that,
for two symmetric quantum wires in resonant tunneling, $V_{J}$ and $V_{H}$
vanish.

\section{Bi-wires without tunneling}

In the following, we will analyze the inelastic Coulomb scattering rate of
electrons in two coupled symmetric quantum wires ($W_{1}=W_{2}=W$) in the
absence of tunneling. As we discussed before, when there is no tunneling
between two quantum wires, $V_{D}=V_{H}=V_{J}=0.$ Only the Coulomb
interactions $V_{A}$, $V_{B}$, and $V_{C}$ contribute to the
electron-electron interaction. Furthermore, the potential $V_{A}$ and $V_{B}$
are responsible for the intrawire interaction and $V_{A}=V_{B}$ due to the
symmetry property of the two wires. The potential $V_{C}$ is responsible for
the interwire Coulomb interaction. If we assume that the two wires have an
identical electron density $n_{1}=n_{2}=n_{e}$, the total electron density
in the system is $N_{e}=2n_{e}$. In this case, the two quantum wires have
the same Fermi level $E_{F}$ so that $\Pi _{11}=\Pi _{22}=\Pi _{0}$.
Therefore, from Eqs. (\ref{diel}) and (\ref{diel1}), we obtain the screened
intrawire Coulomb potential $V_{1111}^{s}=V_{2222}^{s}=V^{s}$ given by 
\begin{equation}
V^{s}=\frac{V_{A}-(V_{A}+V_{C})(V_{A}-V_{C})\Pi _{0}}{[1-(V_{A}+V_{C})\Pi
_{0}][1-(V_{A}-V_{C})\Pi _{0}]}.  \label{vascreened}
\end{equation}
The denominator in the above equation is the determinant of the dielectric
matrix $\det |\epsilon (q,\omega )|$. The equation $\det |\epsilon (q,\omega
)|=0$ yields the plasmon dispersions of the electron gas system. The
plasmons result in singularities in the screened Coulomb potential which are
of the most important contribution to the inelastic Coulomb scattering rate.

According to Eq. (\ref{sigma}), the intrawire scattering rate of the
symmetric bi-wires with identical electron density becomes 
\begin{equation}
\sigma _{n,n}(k)=\frac{1}{2\pi }\int dq\left\{ 
\mathop{\rm Im}%
\left[ V^{s}\left( q,2kq+q^{2}\right) \right] \right\} \left\{ \theta \left(
-2kq-q^{2}\right) -\theta \left( E_{Fn}-k^{2}-q^{2}-2kq\right) \right\} ,
\label{gamma11}
\end{equation}
for $n=1$ and $2,$ where $E_{Fn}=E_{F}-E_{n}$ is the subband Fermi energy.
Notice that $E_{1}=E_{2}$ for two symmetric quantum wires. It is obvious
that $\sigma _{1,1}(k)=\sigma _{2,2}(k)$.\ In the absence of tunneling,
interwire scattering rates $\sigma _{1,2}(k)$ and $\sigma _{2,1}(k)$ are
zero because the transition of an electron from one wire to the other is
impossible. Therefore, we have $\sigma _{1}(k)=\sigma _{1,1}(k)=\sigma
_{2}(k)=$ $\sigma _{2,2}(k)$. But the interwire Coulomb interaction $V_{C}$
influences the collective excitations in the system leading to two different
plasmon modes, i.e., the optical and acoustic modes. Subsequently, it
affects the inelastic-scattering rates. We know that the zeros of the two
parts $1-(V_{A}+V_{C})\Pi _{0}$ and $1-(V_{A}-V_{C})\Pi _{0}$ in the
denominator in equation (\ref{vascreened}) yield the optical and acoustic
plasmon mode dispersions, respectively. To understand better the scattering
mechanism, we show in Fig. 1 the collective excitation dispersion relations
of the two coupled symmetric GaAs quantum wires of width $W=150$ \AA\ with
different barrier widths. In the calculations, we consider the barrier
height $V_{0}=\infty ,$ which does not permit tunneling between the wires.
The plasmon modes in Fig. 1 correspond to the different scattering channels
through which the injected electron can lose energy. We find a higher
(lower) frequency plasmon branch which represents the optical (acoustic)
plasmon mode $\omega _{+}$ ($\omega _{-}$). Intrawire quasiparticle
excitation continuum $QPE$ (shadow region) is also indicated in the figure.
The thin-solid curve is the plasmon dispersion of a single quantum wire with
electron density $n_{e}$. It corresponds to the situation in which the
distance between the two wires is infinity ($W_{b}=\infty $) or $V_{C}=0.$
In this case, the plasmon mode is of dispersion relation $\omega (q)\sim 
\sqrt{n_{e}}q\left| \ln qW\right| ^{1/2}$ at $q\rightarrow 0$.\cite
{sarmagrande} As the distance between the wires decreases, the potential $%
V_{C}$ increases. A finite $V_{C}$\ leads to a gap between the two plasmon
modes. When the two wires are close enough, the acoustic mode develops a
linear wave vector dependence. For $q\rightarrow 0$, $\omega _{-}(q)=vq$
with $v=[v_{F}+4V_{-}(q=0)/\pi ]$ where $v_{F}$ is the Fermi velocity and $%
V_{-}(q)=V_{A}(q)-V_{C}(q)$, whereas the optical plasmon still keeps its
well-know 1D dispersion relation, $\omega _{+}(q)\sim \sqrt{N_{e}}q\left|
\ln qW\right| ^{1/2}$.\cite{hwang,prlnosso} Notice that, the interwire
Coulomb interaction $V_{C}$, depending on the distance between the two
wires, is responsible for the behavior of the wavevector dependence\ of the
acoustic mode. As we will see, this affects significantly the inelastic
Coulomb scattering rate due to the acoustic plasmons.

Fig. 2 shows the numerical results of inelastic plasmon scattering rate in
the coupled wires corresponding to Fig.1(a) with a very small broadening
constant $\gamma =10^{-4}$ meV. We observe two scattering peaks. The lower
(higher) one is due to the acoustic (optical) plasmon scattering. The abrupt
increase of the scattering rate at threshold electron momenta $k_{c}^{-}$
and $k_{c}^{+}$\ correspond to the onset of the scattering of the acoustic
and optical plasmon modes, respectively. The higher scattering peak due to
the optical plasmon mode is always divergent at the onset $k=k_{c}^{+}$ and $%
\sigma _{1,1}(k)\propto (k-k_{c}^{+})^{-1/2},$ similarly to that in the
single wire. But the behavior of the lower scattering peak is dependent on
the distance between the two wires which is directly related to the
dispersion relation of the acoustic plasmon mode at small $q$. For small $%
W_{b},$\ the acoustic mode is of a linear wavevector dependence leading to a
finite scattering rate at the onset $k=k_{c}^{-}$. With increasing $W_{b}$,
the acoustic mode loses its linear $q$ dependence resulting in the
divergency at the onset of the scattering. In order to clarify such a
behavior, we show in the inset the energy- vs momentum-loss curve 
\begin{equation}
\omega _{k}(q)=2kq-q^{2}
\end{equation}
for $k=k_{c}^{+}\simeq 2.13\times 10^{6}$ cm$^{-1}$ (thin-solid curve) and $%
k_{c}^{-}\simeq 1.65\times 10^{6}$ cm$^{-1}$ (thin-dashed curve) in the
system with $W_{b}=30$ \AA . Along these curves, the momentum and energy
conservations are obeyed and the electron relaxation is allowed. The
dispersions of the optical and acoustic plasmon modes $\omega _{+}(q)$ and $%
\omega _{-}(q)$ are also given by thick long-dashed curves in the same
figure. For $k=k_{c}^{+}$ ($k_{c}^{-}$), the thin-solid (thin-dashed) curve\
intersects the optical (acoustic) mode dispersion curve at $q=q_{c}^{+}$ $%
(q_{c}^{-})$. This means that the injected electron with momentum $k_{c}^{+}$
($k_{c}^{-}$) can emit one optical (acoustic) plasmon of frequency $\omega
_{+}(q_{c}^{+})$ ($\omega _{-}(q_{c}^{-})$). Notice that, the slopes of the
curves $\omega _{k_{c}^{+}}(q)$ ($\omega _{k_{c}^{-}}(q)$) and $\omega
_{+}(q)$ ($\omega _{-}(q)$) are equal at $q=q_{c}^{+}$ ($q_{c}^{-}$). For
the optical plasmon mode, the intersection always occurs at finite $q_{c}^{+}
$ because the optical plasmon goes as $\omega _{+}(q)\sim q\left| \ln
qW\right| ^{1/2}$ for small $q$. The divergency due to the optical plasmon
scattering is similar to that in the single quantum wire\cite{sarmagrande}
which is resulted from the coupling of the initial and final states via
plasmon emission at $k=k_{c}^{+}$. However, for the acoustic plasmon mode
with linear $q$-dependence, $q_{c}^{-}=0$ because $\omega _{k}(q)\rightarrow
2kq$ at $q\rightarrow 0$. In this case, one can obtain $k_{c}^{-}=v/2$. Due
to the fact that the plasmon mode is of vanished oscillator strength at $q=0$%
, the emission of the acoustic plasmon of the wavevector $q=q_{c}^{-}$
cannot produce divergency in the inelastic-scattering rate. With increasing
the distance between the two wires, the acoustic plasmon mode loses its
linear $q$-dependence and approaches to the dispersion of the optical
plasmon mode. Consequently, the $q_{c}^{-}$ becomes finite and the
scattering rate is divergent at the threshold momentum $k_{c}^{-}$.\ In Fig.
3, we show the scattering rates in the same structures as in Fig. 2 but with
higher electron density $n_{e}=10^{6}$cm$^{-1}$. We see that, in the systems
of higher electron density, the scattering threshold shift to larger
momentum and the scattering is enhanced.

In Fig. 2 and 3, we have not shown the inelastic-scattering rate due to
virtual emission of quasiparticles which would occur below the threshold
wavevector. It is known that, in a one-subband quantum wire, the
contribution of the quasiparticle excitations to the inelastic Coulomb
scattering rate is completely suppressed due to the restrictions of the
energy and momentum conservations. Consequently, the scattering rate is zero
until the onset of the plasmon scattering at a threshold $k_{c}>k_{F}$. \cite
{sarmagrande} The quasiparticle excitations contribute to the inelastic
scattering only when the level broadening is introduced. These contributions
are negligible when the broadening constant is small. Although, in the
present case, we are dealing with two coupled quantum wires, the Coulomb
interaction does not influence the quasiparticle excitations as well as
their contributions to the inelastic scattering.

As far as the effect of the phenomenological broadening constant $\gamma $
is concerned, we show in Fig. 4 the dependence of the inelastic-scattering
rate for different $\gamma $. Finite broadening values of $\gamma $ in the
system give rise to the breaking of translational invariance due to the
presence of impurity. This fact is responsible for relaxing the momentum
conservation permitting inelastic scattering via quasiparticle and plasmon
excitations for $k<k_{c}^{\pm }.$ We show such a contribution in the inset
of Fig. 4. For $k=k_{F}\simeq 1.6\times 10^{6}$ cm$^{-1}$, conservation of
energy and momentum does not permit opening of any excitation channels. This
means that the injected-electron has infinite lifetime at the Fermi surface
which has been restored by impurity effects.

\section{Weak tunneling effects}

In this remaining section, we are going to discuss the effect of weak
tunneling on the inelastic Coulomb scattering rates in two coupled symmetric
quantum wires as have been shown in the previous section. When the tunneling
occurs, a energy gap $\Delta _{SAS}$ opens up between the two lowest
subbands which have symmetric and antisymmetric wavefunctions in the $y$%
-direction about the center of the barrier. In this case, only the subband
index is a good quantum number. As we have seen in section II, $V_{J}$ and $%
V_{H}$ vanish in two symmetric quantum wires in resonant tunneling. But, $%
V_{D}$ is finite and responsible for the tunneling effects on the Coulomb
scattering. In weak resonant tunneling condition, one finds $V_{A}\simeq
V_{B}\simeq V_{C}\simeq U$. After some algebra, we obtain 
\begin{equation}
V_{1111}^{s}=\frac{1+U(\Pi _{11}-\Pi _{22})}{1-U\left( \Pi _{11}+\Pi
_{22}\right) }U,  \label{v1}
\end{equation}
\begin{equation}
V_{2222}^{s}=\frac{1-U(\Pi _{11}-\Pi _{22})}{1-U\left( \Pi _{11}+\Pi
_{22}\right) }U,  \label{v2}
\end{equation}
\begin{equation}
V_{1221}^{s}=\frac{1+V_{D}(\Pi _{12}-\Pi _{21})}{1-V_{D}\left( \Pi _{12}+\Pi
_{21}\right) }V_{D},  \label{v12}
\end{equation}
and 
\begin{equation}
V_{2112}^{s}=\frac{1-V_{D}(\Pi _{12}-\Pi _{21})}{1-V_{D}\left( \Pi _{12}+\Pi
_{21}\right) }V_{D}.  \label{v21}
\end{equation}
From the above equations and equation (\ref{sigma}), we can obtain the
inelastic Coulomb scattering rates in the presence of tunneling. We also
notice that the zeros of the denominators in equations (\ref{v1}) and (\ref
{v2}) yield the optical plasmon dispersion and those in equations (\ref{v12}%
) and (\ref{v21}) yield the acoustic plasmon dispersion. It indicates that
the optical plasmons only contribute to the intrasubband scattering $\sigma
_{11}$\ and $\sigma _{22}$, and the acoustic plasmons to the intersubband
scattering $\sigma _{12}$\ and $\sigma _{21}$.

We consider two coupled GaAs/Al$_{0.3}$Ga$_{0.7}$As ($V_{b}=228$ meV)
quantum wires of widths $W_{1}=W_{2}=150$ \AA\ separated by a barrier of $%
W_{b}=70$ \AA . In this case, we find $\Delta _{SAS}=0.14$ meV indicating a
very weak resonant tunneling. We show in Fig. 5(a) both inter- and
intra-subband scattering rates. The intrasubband scattering rates $\sigma
_{11}$\ and $\sigma _{22},$ induced by the emission of the optical plasmons,
is very similar to that in the absence of tunneling. It is also not
difficult to understand that $\sigma _{11}\simeq \sigma _{22}$ because,
above the threshold of the optical plasmon emission, the plasmon frequency
is much larger than $\Delta _{SAS}$ and consequently, $\Pi _{11}\simeq \Pi
_{22}.$ On the other hand, the tunneling introduces the intersubband
scattering rates $\sigma _{12}$\ and $\sigma _{21}$ \ and modifies strongly
the mechanism of the acoustic plasmon emission. In order to clarify such
results, we plot the corresponding acoustic plasmon dispersion relation\ in
thick-dashed curve in Fig. 5(b). The acoustic mode develops a plasmon gap at
zero $q$ due to the tunneling effect.\cite{prlnosso} The thin lines indicate
the intersubband energy- vs momentum-loss curves at the onsets of the
acoustic plasmon scattering. They are determined by conservations of energy
and momentum, given by 
\begin{equation}
\omega _{k}^{12}(q)=2qk-q^{2}-\Delta _{SAS}
\end{equation}
for $k=k_{c}^{12}$ (thin-dashed curve), and 
\begin{equation}
\omega _{k}^{21}(q)=2qk-q^{2}+\Delta _{SAS}
\end{equation}
for $k=k_{c}^{21}$ (thin-dotted curve), where $k_{c}^{12}$ and $k_{c}^{21}$
are threshold wavevectors above which the injected electron can be
transferred to a different subband by emitting an acoustic plasmon. The $%
\omega _{k}^{21}(q)$ (thin-dotted curve) intersects the acoustic plasmon
dispersion at small wavevector $q=q_{c}^{21}\simeq 0.05\times 10^{6}$cm$%
^{-1} $. The scattering process is similar to the acoustic plasmon
scattering in the absence of tunneling as we discussed in the previous
section. But now, the acoustic plasmon mode is of finite frequency with also
a finite oscillator strength at $q\rightarrow 0$ resulting in a small
divergency at $k_{c}^{21}$. On the other hand, the intersection between the $%
\omega _{k}^{12}(q)$ (thin-dashed curve) and the acoustic plasmon dispersion
occurs at a quite larger wavevector $q=q_{c}^{12}\simeq 0.18\times 10^{6}$ cm%
$^{-1}$. The scattering mechanism is more similar to that of the
intrasubband scattering and produces a pronounceable divergence at $%
k_{c}^{12}$.

Finally, we would like to show the tunneling effects on the total inelastic
Coulomb scattering rates $\sigma _{n}(k)=\sum_{n^{\prime }}\sigma
_{n,n^{\prime }}(k)$. Fig. 6 gives the total scattering rates in (a) the
absence and (b) the presence of tunneling between two quantum wires with $%
W=150$ \AA\ and $W_{b}=70$ \AA . We observe that a weak resonant tunneling
does not influence much the optical-plasmon scattering, but it does affect
strongly acoustic-plasmon scattering. The acoustic-plasmon scattering for
the injected electron in the lowest subband is enhanced significantly and a
quite strong scattering peak appears. For the injected electron in the
second subband, tunneling introduces a small divergency in the scattering
rate and\ shifts the scattering threshold to the lower wavevector.

\section{Summary}

We have calculated the inelastic Coulomb scattering rates of two coupled Q1D
electron gas systems within the GW approximation. The screened Coulomb
potential was obtained within the RPA. The Coulomb interaction between the
two quantum wires leads to the optical and acoustic plasmon modes and,
consequently, two scattering peaks appear due to the scattering of the two
modes. We found that the scattering of the optical plasmons in two coupled
quantum wires is very similar to the plasmon scattering in a single wire
because both plasmon modes have similar dispersion relations at small $q$.\
The scattering rate is divergent at the onset of the optical plasmon
scattering. However, the acoustic plasmon mode does not produce such a
divergency when it is of a linear $q$-dependence at small $q$. \ This
happens when two wires are close enough. Furthermore, we studied the
tunneling effects on the inelastic scattering. A weak resonant tunneling was
introduced between the wires. Such a tunneling lifts the degeneracy of the
two subbands originated from two quantum wires and also produces a small
plasmon gap on the acoustic mode at $q=0.$ Moreover, intrersubband
scattering appears. We show that, in this case, the optical plasmons are
only responsible for the intrasubband scattering and the acoustic plasmons
are for the intersubband scattering. A weak tunneling enhances significantly
acoustic plasmon scattering for an injected electron in the lowest subband.

\section{Acknowledgments}

This work is supported by FAPESP and CNPq, Brazil.

\begin{figure}[tbp]
\caption{Dispersions of the collective excitations of two coupled quantum
wires of (a) $n_e=0.5\times 10^6$ cm$^{-1}$ and (b) $n_e=10^6$ cm$^{-1}$
with $W_1=W_2=150$ \AA\ and $W_b=300$ \AA\ (dotted curves), 70 \AA\ (dashed
curves), and 30 \AA (long-dashed curves). Plasmon mode of the corresponding
single wire ($W_b=\infty $) is presented by the thin-solid curves. The
shadow areas indicate the quasiparticle continua.}
\end{figure}

\begin{figure}[tbp]
\caption{The inelastic Coulomb scattering rates correspondig to Fig. 1(a)
with $n_e=0.5\times 10^6$ cm$^{-1}$. Inset shows the acoustic and optical
modes (thick-dashed lines) for $W_b=30$ \AA, and the intrawire energy- vs
momentum-loss curves at the onset of the optical (thin-solid line) and
acoustic (thin-dashed line) plasmon scattering. }
\end{figure}

\begin{figure}[tbp]
\caption{The same as Fig. 2 but now with $n_e=10^6$ cm$^{-1}$.}
\end{figure}

\begin{figure}[tbp]
\caption{The inelastic-scattering rate in coupled wires of $n_e=10^6$ cm$%
^{-1}$, $W_1=W_2=150$ \AA, and $W_b=70$ \AA\ for different values of the
broadening constant $\protect\gamma=10^{-4}$ (thin line), $0.01$ (dotted
line), $0.1$ (dashed line), and $1$ meV (long-dashed line).}
\end{figure}

\begin{figure}[tbp]
\caption{(a) The intra- and inter-subband inelastic-sacattering rates in two
coupled GaAs/Al$_{0.3}$Ga$_{0.7}$As quantum wires with tunneling. $%
W_1=W_2=150$ \AA, $W_b=70$ \AA\ and $N_e=10^6$ cm$^{-1}$. The solid curves
present $\protect\sigma _{1,1}(k)$ and $\protect\sigma _{2,2}(k)$. The
dashed and dotted curves present $\protect\sigma _{1,2}(k)$ and $\protect%
\sigma _{2,1}(k)$, respectively. (b) The acoustic plasmon dispersion $%
\protect\omega_{-}(q)$ (thick-dashed curve) in the system. The thin-dashed
line indicates the $\protect\omega _{k}^{12}(q)$ curve for $k_{c}^{12}\simeq
1.79$ $\times 10^{6}$ cm$^{-1}$ and the thin-dotted line indicates the $%
\protect\omega_{k}^{21}(q)$ curve for $k_{c}^{21}\simeq 1.59$ $\times 10^{6} 
$ cm$^{-1}$. $n_1=0.51\times 10^6$ cm$^{-1}$ and $n_2=0.49\times 10^6$ cm$%
^{-1}$.}
\end{figure}

\begin{figure}[tbp]
\caption{ Total inelastic-scattering rate $\protect\sigma _{n}(k)$ of the
bi-wire system (a) without ($V_{0}=\infty $) and (b) with tunneling ($%
V_{0}=228$ meV). $W_1=W_2=150$ \AA, $W_b=70$ \AA\ and $N_e=10^6$ cm$^{-1}$. }
\end{figure}

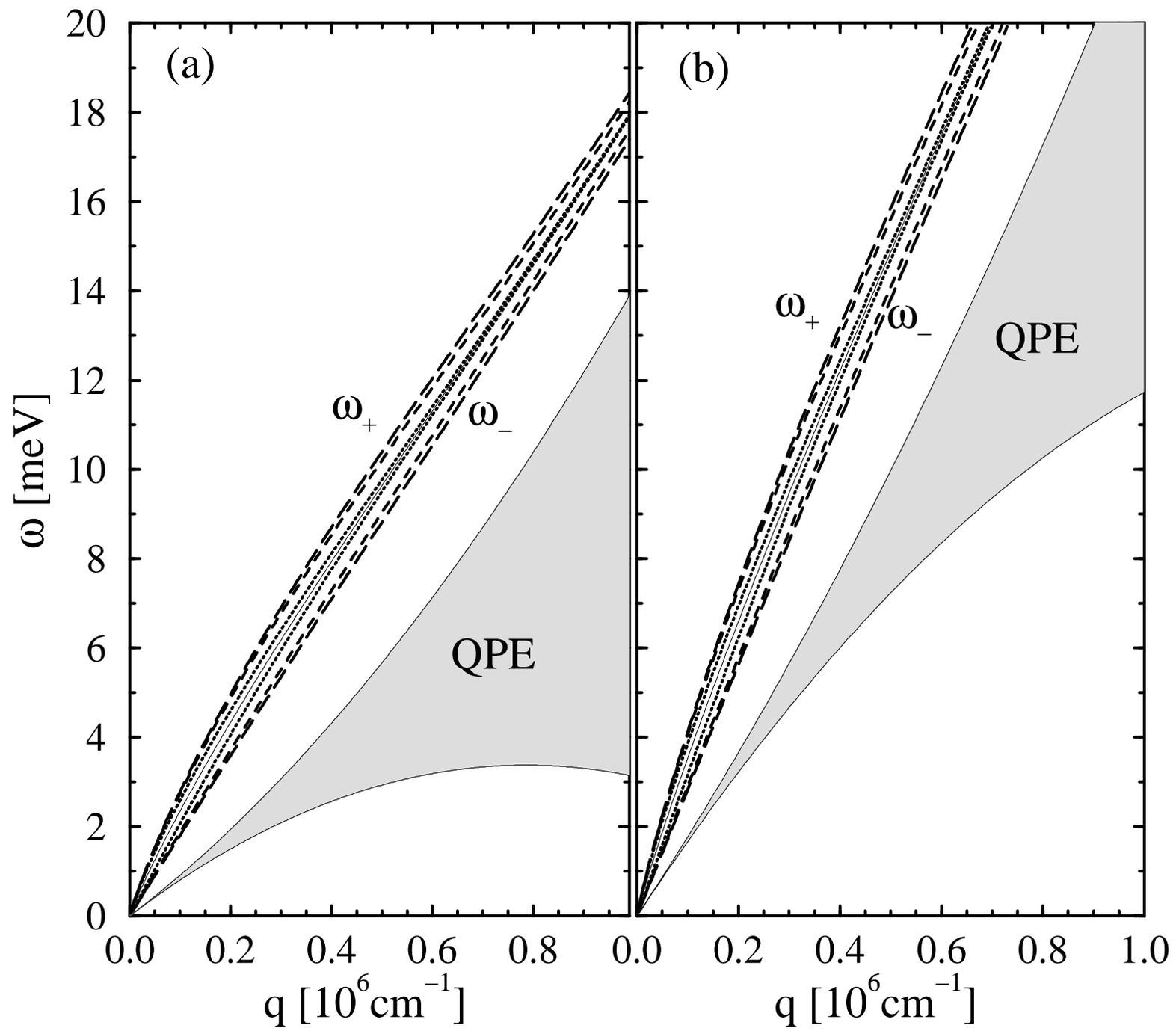
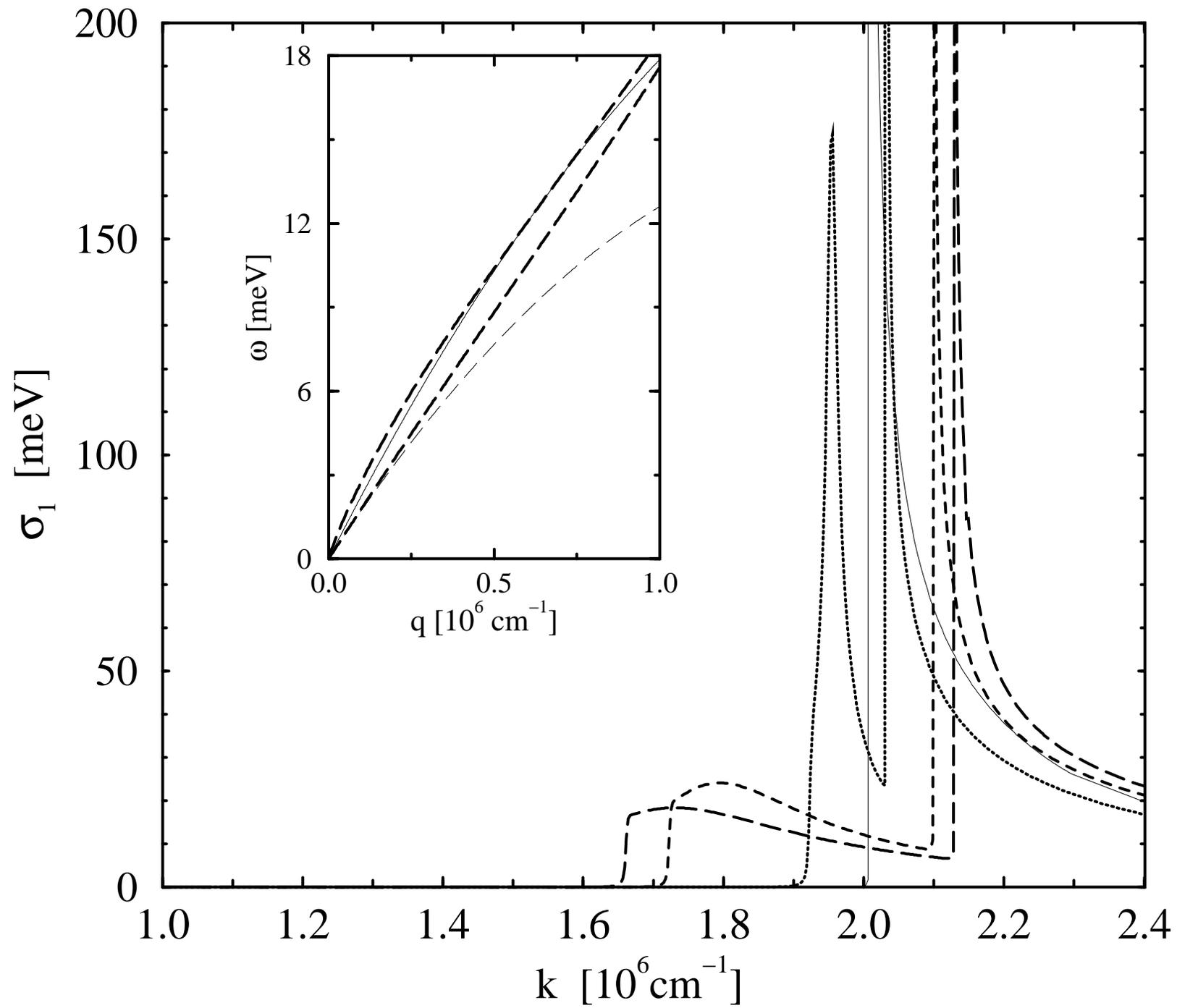
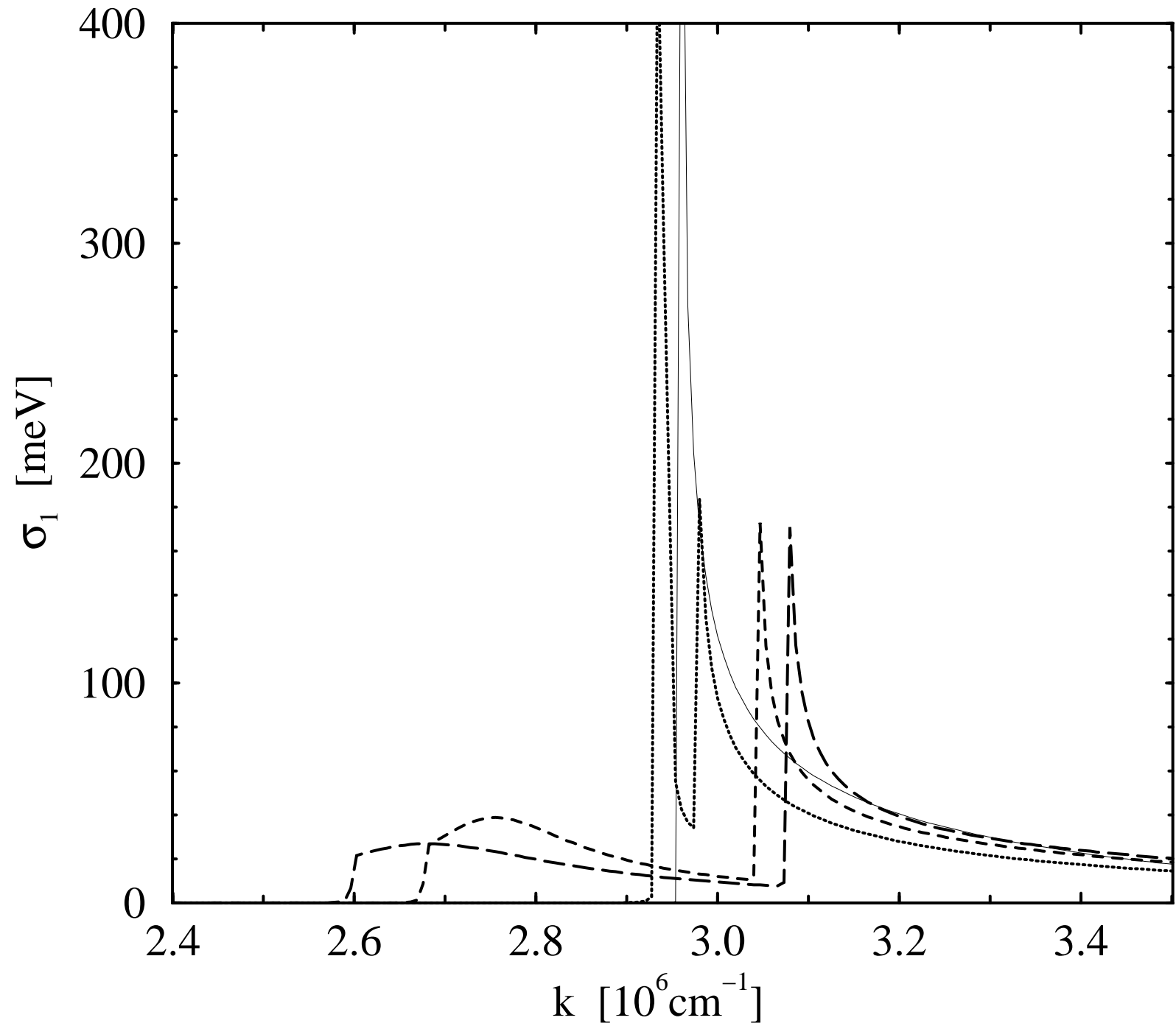
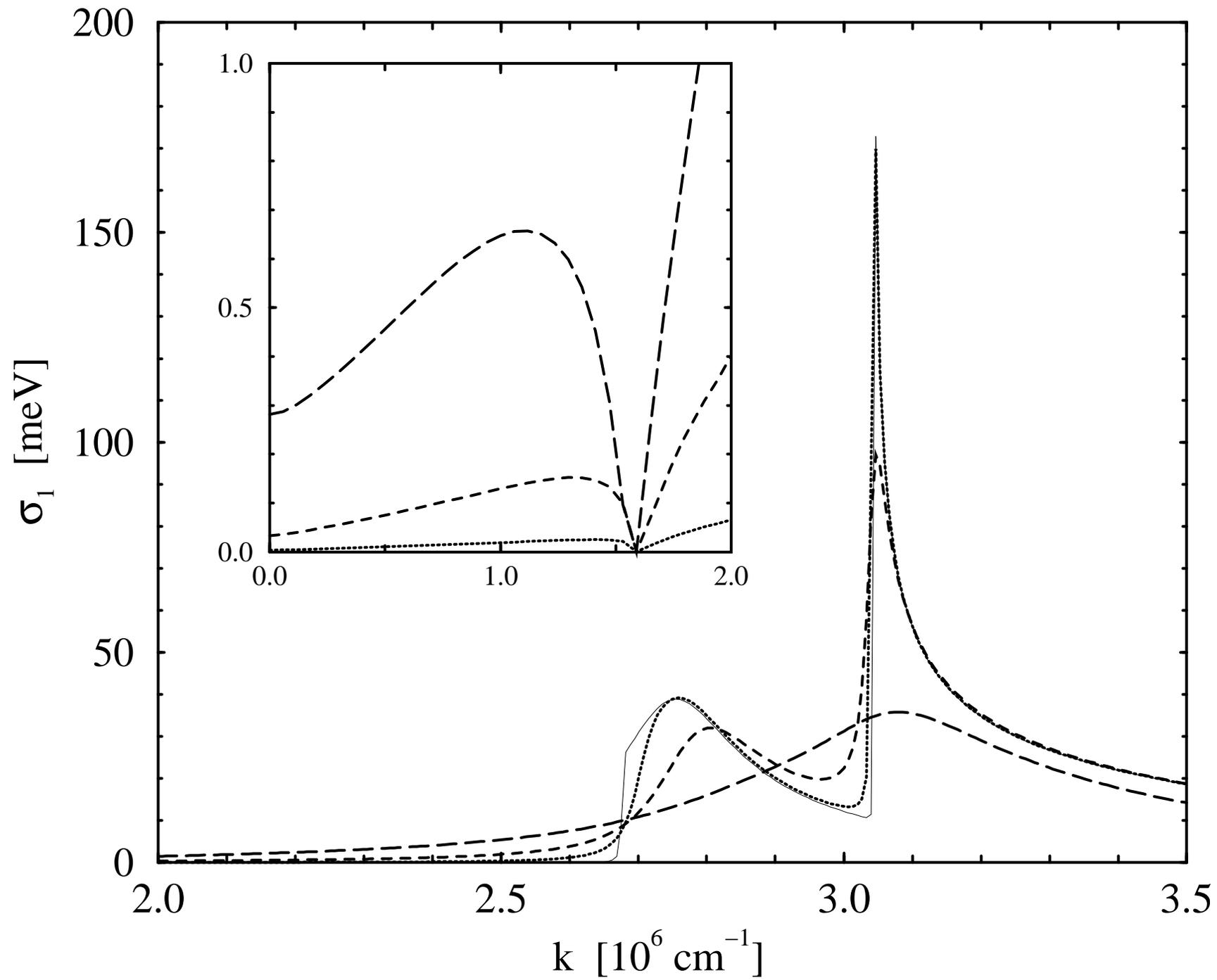
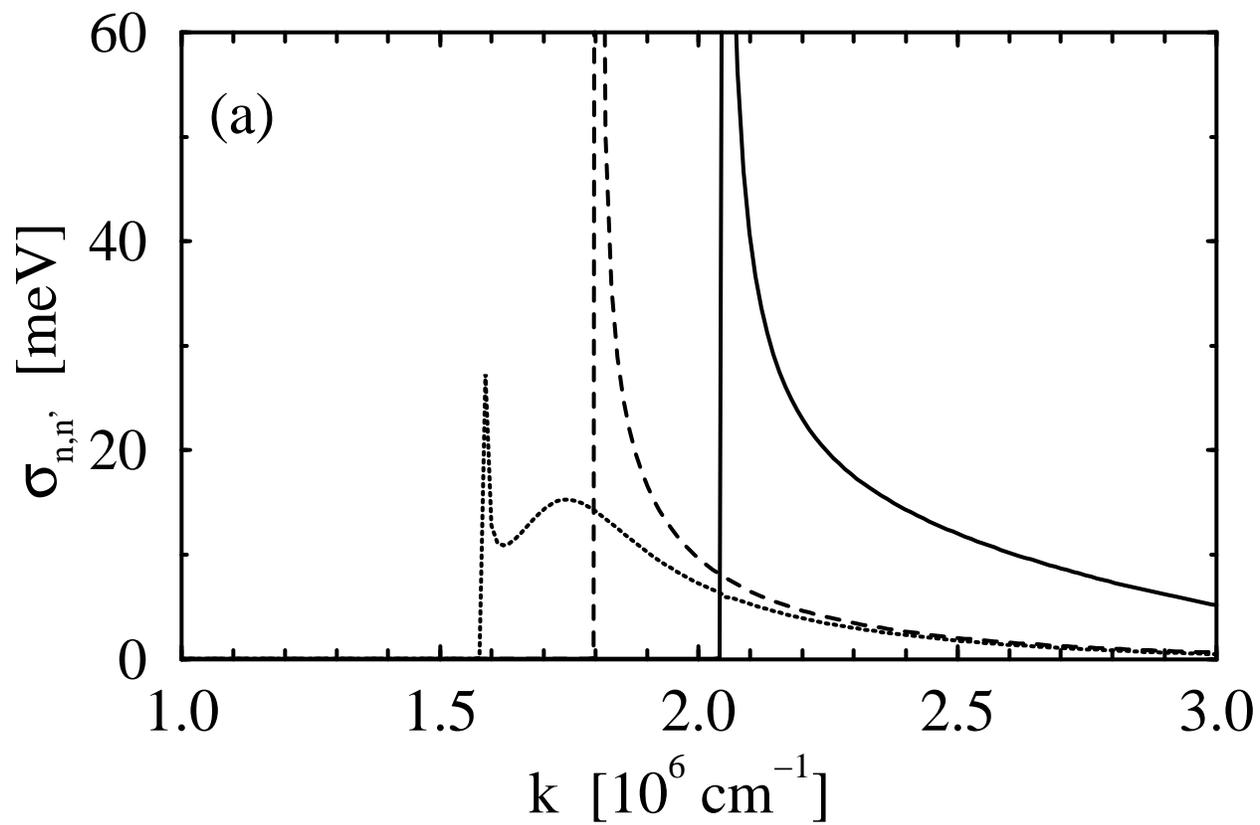
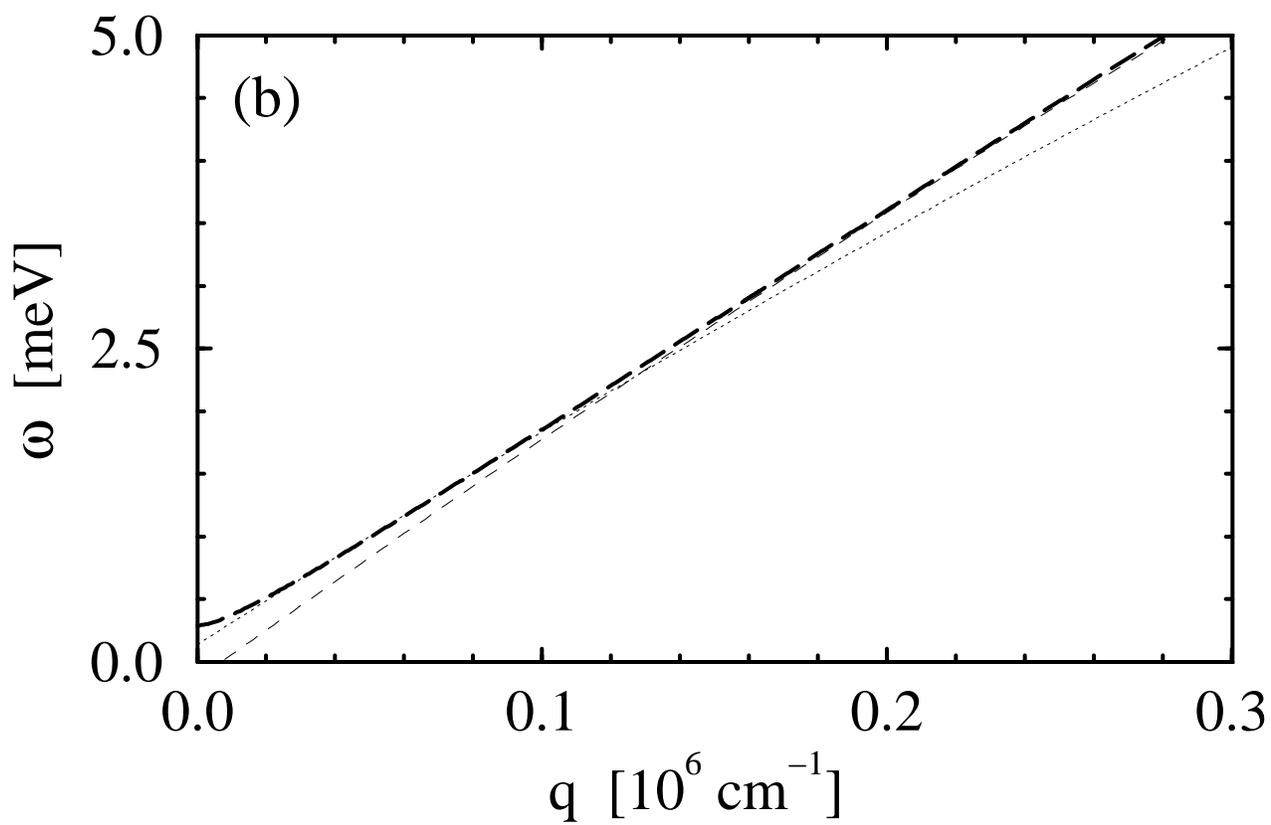
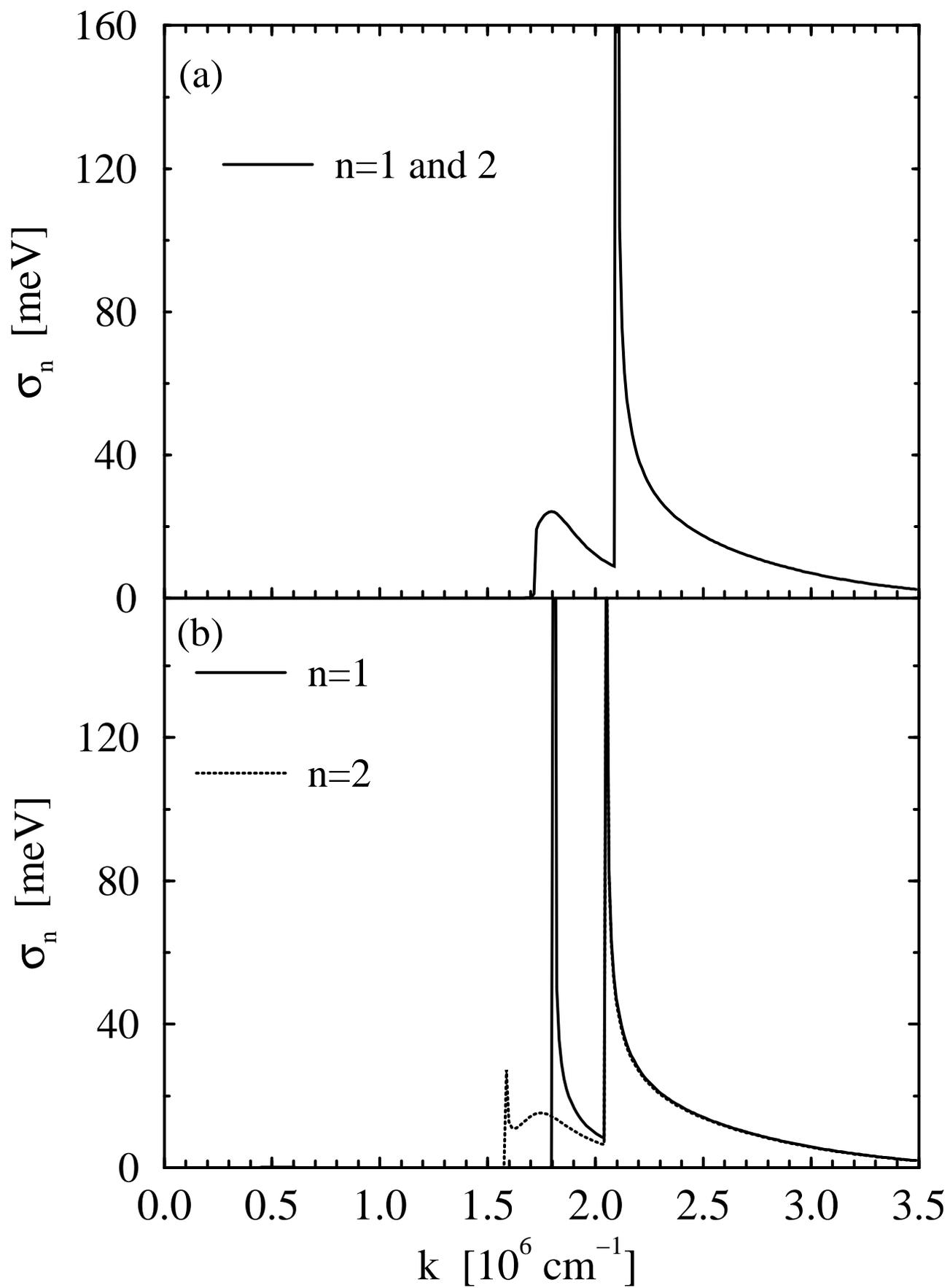
\end{document}